# Polar Solvent Vapour Induced Photo-Physical Properties on the Solid-State Organometallic Halide Perovskite Nanocrystals

Sang-Hyun Chin,[†][ab] Jin Woo Choi,[†][a] Hee Chul Woo,[a] Jong H. Kim,[c] Hong Seok Lee*[b] and Chang-Lyoul Lee*[a]

Advanced Photonics Research Institute (APRI), Gwangju Institute of Science and Technology (GIST), Gwangju, 61005, Republic of Korea. E-mail: vsepr@gist.ac.kr.

Department of Physics, Research Institute of Physics and Chemistry (RINPAC), Chonbuk National University, Jeonju, 54896, Republic of Korea. E-mail: hslee1@jbnu.ac.kr

Department of Molecular Science and Technology, Ajou University, Suwon, 16499, Republic of Korea.

† These authors contributed equally.

## ABSTRACT

The optoelectric applications, light emitters and photovoltaics, with organometallic halide perovskite materials have been decided by dynamics of charge carriers in grains on the polycrystalline thin film s. Especially, polar solvent vapour recrystallizes perovskites and changes their grain sizes. In this paper, various polar solvents were evaluated with density functional theory calculation. And a stociometric treatment with these solvents which enhanced photoluminescence intensity was established. The properties of solvents which caused dramatic photoluminescence enhancement and quantum confinement effect were discussed.

## Introduction

All-solid organometallic halide perovskites $ABX_3$ (A = $CH_3NH_3^+$, $CH_2(NH_3)_2^+$, B = $Pb^{2+}$, X = $Cl^-$, $Br^-$, $I^-$) solar cell have shown significant progress on power conversion efficiency (PCE) after emergence of $MAPbI_3$ perovskites dye-sensitized solar cells by A. Kojima et al in 2009.[1-7] Due to superb optoelectronic properties of $MAPbX_3$, for instance narrow full width at half maximum (FWHM) around 20 nm, high photoluminescence quantum yield (PLQY) over 80 % in thin films and band gap tunability with halide anion exchange, their potential area of application is extended to light-emitters.[8-10] In light emitting diode (LED) application, forming small grains which confine exciton is proper to radiative recombination as a result of reduced diffusion length.[11] For this reason, recent researches about organometallic halide perovskite (OHP) LED applications are mainly focused on methods for reducing perovskite grain size. For instance, anti-solvent engineering with or without additive attributed to hindrance of large grain growth and solvent-vacuum drying induced fast solvent evaporation for small grain formation.[11,12] Except size effect, OHP active layer with low defect density is indispensable condition for high performance LED fabrication. Unfortunately, moisture which is abundant in ambient air causes decomposition of OHP materials easily and produces sub-grain defects. Extremely high polarity of $H_2O$ molecule is the most relevant factor of solubility for the OHPs. Moisture strongly interacts with organometallic cations due to its high polarity and produces defect after $MAPbX_3$ decomposition.[13]

However, there were few reports that humid environment assisted conversion of precursors to OHPs and enhanced optoelectronic device performances.[14-16] The precedent research of LED application which employed simple exposure on the humid environment without precise treatment reported 6 times higher electroluminescence intensity compared to fabrication in $N_2$ environment due to reduced defect density.[16] According to these works, taking advantage to humidity exposure as a polar solvent treatment seems to be a promising procedure with high potential.

In spite of these importance of humidity to optoelectronic applications of OHPs, there are only few reports exist.[17] Therefore, studies on various solvents with high polarity and vapour pressure which are applicable to OHP treatments is required. Herein, we report a novel reliable method to control the grain size of OHP films for optimizing light emission properties with the aid of polar solvent vapour annealing (SVA).

## Result and Discussion

For OHP films fabricated by spin coating process, thermal annealing is often used as post treatment to evaporate remaining solvents and facilitate crystallization process. Therefore, the temperature and the rate of heating are of great importance. On the other hands, for OHP films fabricated by VASP, there are remaining PbBr$_2$ and MABr that were not fully crystallized into MAPbBr$_3$ perovskite film. Furthermore, not much of solvents is remained in as-fabricated OHP films. To this end, direct thermal annealing right after VASP may not be an efficient manner to crystalize the films.

The basic concept of SVA treatment is to promote conversion of such remaining PbBr$_2$ and MABr into MAPbBr$_3$ perovskite layer and recrystallize already fabricated OHP films, thereby improve crystallinity and control the grain size simultaneously. During such treatment, both unreacted precursors and synthesized perovskites film undergo recrystallization process which is highly influenced by solvent polarity and vapour pressure. Therefore, the crystallinity and the grain size of OHP film are determined by adjusting both polarity and vapour pressure of the solvent.

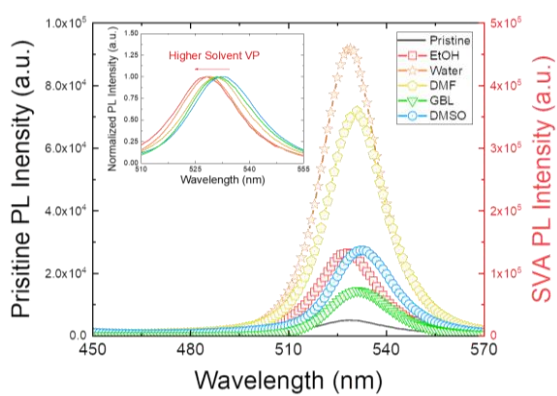

**[Fig. 1]** Photoluminescence spectra before and after SVA with various solvents. And normalized photoluminescence spectra after SVA with varied peaks due to their vapour pressures (inset).

Fig. 1 plots the photoluminescence of MAPbBr$_3$ thin films treated with various polar solvents such as ethanol (EtOH), water (H$_2$O), dimethylformamide (DMF), gamma-butylactone (GBL) and dimethylsulfoxide (DMSO). The detailed VASP process and the SVA treatment are described in Supplementary Information† with schematics.[18] As previously mentioned, the grain size of the OHP films are influenced by two important factors; solvent polarity and solvent vapour pressure. The values of solvent polarity and vapour pressure are listed in table 1. Among various solvents, ethanol shows the highest solvent polarity and vapour pressure while H$_2$O shows the second highest. In general, higher solvent polarity enables better solubility of ionic compounds.

**[Table 1]** Properties of polar solvent and PL peak wavelengths

|  | EtOH | Water | DMF | GBL | DMSO |
|---|---|---|---|---|---|
| VP (mmHg) at 20 °C | 44.6 | 17.7 | 2.7 | 1.5 | 0.42 |
| Solvent Polarity (kJ/mol) | 51.9 | 63.1 | 43.8 | 44.3 | 45.0 |
| Peak wavelength (nm) | 528 | 529 | 531 | 531 | 533 |

Therefore, PL intensity may increase for OHP films treated by solvent with higher polarity because of additional perovskite layer formed by crystallization of remaining precursors. In this regard, the strongest PL intensity is expected for ethanol and H$_2$O compared to conventional solvent such as DMF, GBL and DMSO.

However, PbBr$_2$ is known to be not soluble in alcohol. Therefore, despite its high polarity, ethanol cannot effectively PbBr$_2$ and OHP films treated by H$_2$O shows the strongest PL intensity. Meanwhile, the highest vapour pressure of ethanol still enables recrystallization of already fabricated MAPbBr$_3$ films because OHP is soluble in alcohol. Higher vapour pressure implies faster solvent evaporation during recrystallization process and this results in smaller grain size. The normalized PL spectra are represented in Fig. 1 inset, and the peak position gradually moves to lower wavelength with increasing solvent vapour pressure. Small grains not only guarantee higher exciton binding energy but also cause quantum confinement effect, leading to widening of energy band gap, and thus blue shift in PL.[19,20] Therefore, OHP film, treated by ethanol shows the most blue-shifted PL that is originated from the smallest grain.

Consequently, the strongest PL intensity was observed for MAPbBr$_3$ film treated not by ethanol but by H$_2$O due to both high solubility toward remaining precursors and vapour pressure. Furthermore, this recrystallization process is observed to be reversible which means that OHP films repeat alternating their phase from liquid-solid to nanocrystals under humid and dry condition respectively. Supplementary Information demonstrates this reversible recrystallization process very well. MAPbBr$_3$ film is sealed in a petri dish with H$_2$O droplets and put onto UV lamp to monitor the PL in real time. H$_2$O droplets made petri dish be filled with water vapour and PL intensity was decreased because perovskite changed its phase to liquid-solid. When the water vapour is released by opening petri dish, the liquid-solid perovskite now recrystallizes and turns to solid nanocrystals showing strong PL intensity due to increased crystallinity with high exciton binding energy originated from small grain size.

To investigate further the reversible recrystallization process, the interaction distance between various solvents and methylammonium ion (CH$_3$NH$_3^+$) is estimated by theoretical calculation based on density functional theory (DFT). The table with calculated values are attached on Fig. S2 of Supplementary Information† and all the theoretical calculations were performed by DMol3 program in the Materials Studio 4.4 package which is the quantum mechanical code.[21] Among various solvent H$_2$O shows the shortest interaction distance, 1.18 Å.

We assume that longer interaction distances of other solvent allow better reversible recrystallization process

due to reduced interaction strength between the solvents and $CH_3NH_3^+$ ions.[22]

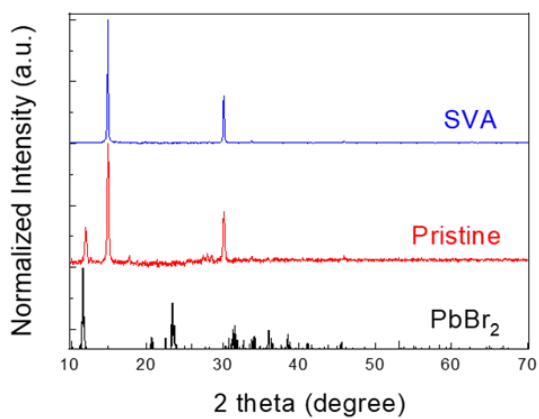

**[Fig. 2] Normalized XRD patterns of thin films ($PbBr_2$, pristine perovskite, after SVA treatment, respectively).**

The effect of SVA treatment was further confirmed by X-ray diffraction (XRD) (Fig. 2). Typical XRD peak originated from (100) plane of $MAPbBr_3$ perovskite appears at 15°. The intensity of this peak increased almost 6 times higher after SVA treatment with $H_2O$. In addition, the peak at 12° originated from $PbBr_2$ was significantly decreased after SVA.[23] This result verifies greatly improved crystallinity of $MAPbBr_3$ perovskite and additional crystallization of remaining MABr and $PbBr_2$. Fig. 3(a) shows the evolution of PL as a function of solvent vapour exposure time in $MAPbBr_3$ film. As expected, in the beginning of SVA treatment, blue shift of PL peak and enhanced luminance was observed. In this stage, $H_2O$ might dissolved already fabricated perovskites and moved through pin-holes blending remaining precursors. Then the solvent vapour completed the reaction of remaining precursors, thereby produced additional perovskite layer which contributes to enhancement of PL intensity. Further exposure to the solvent after when the PL intensity reaches its maximum results in red shift of PL peak with decrease in PL intensity.

This is due to the excess of solvent that turns perovskites again to liquid-solid phase. The deconvolution of PL spectrum during SVA treatment has shown two separated peaks at 480 nm (2.58 eV) and 510 nm (2.43 eV) (Fig. S2). [R15] The higher emission energy must be due to quantum confinement effect which can be described by following equation where $E_{QCE}$ (eV) is

$$E_{QCE} = 2.39 + \frac{12}{D^2}$$

the energy gap of thin films taking account of quantum confinement effect and $D$ (nm) is the mean grain size of the thin films. Then the wavelength of 480 nm and 510 nm correspond to grain size of 8 nm and 20 nm respectively which is confirmed by SEM image (Fig. 3(b-e)).[20] Grains with micrometre-size (Fig. 3(b)) is responsible for PL at 530 nm while those with 20 nm (Fig. 3(c)) is responsible for PL at 510 nm. In addition, as the solvent vapour first reached to the already fabricated perovskite grains, recrystallized grains with 20 nm-diameter are on the 8 nm-diameter grains. This result is not in consistent with previous report from other groups that $H_2O$ does not affect the band gap of OHP thin films.[17] Fig. 3(d) shows reduced pin-holes in thin film for prolonged exposure time with is equivalent to SVA II. After the longest SVA (SVA III) on this experiment, grains on Fig. 3 (e) exceeded 5 micrometres. This treatment need optimization for light emitting application in terms of trade-off between precursor conversion and exciton diffusion enhancement owing to recrystallized bigger grains.

Exposure time dependent time-resolved photoluminescence (TRPL) was also recorded for the $MAPbBr_3$ films and four samples are prepared. The carrier lifetime (τ) was fitted by a bi-exponential function as expressed below.[11,24,25]

$$A = A_1 e^{-\frac{t}{\tau_1}} + A_2 e^{-\frac{t}{\tau_2}}$$

The average carrier lifetimes were gradually increased until SVA II measurement and decreased at SVA III with a similar tendency to PL intensity. Usually, OHP materials are filled with high population of defects acting as non-radiative trap sites for carriers.

**[Fig. 3] (a) PL spectra shift of OHP thin film by moisture exposed time and SEM images of perovskite grains depend on solvent vapour exposure time. Magnified 30k times for (b) Pristine, 100k times for (c) SVA I and (d) SVA II, 5k times for (e) SVA III, respectively**

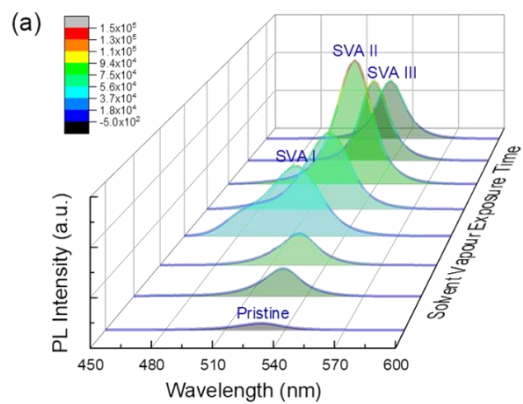

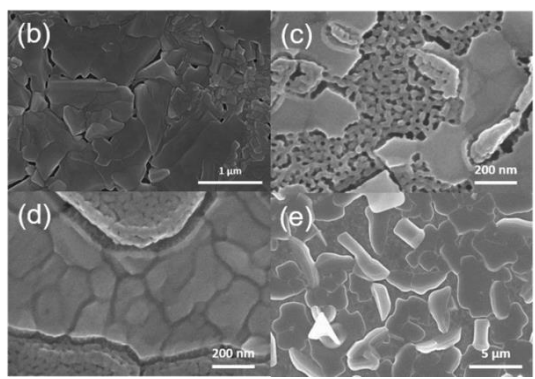

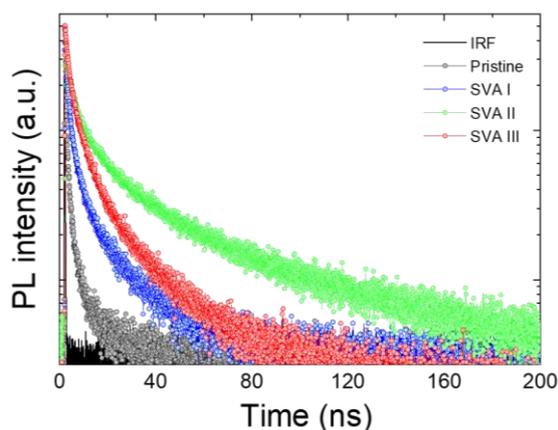

**[Fig. 4]** TRPL spectra of pristine and treated MAPbBr$_3$ thin films excited at 405 nm and profiled at 530 nm.

Due to their relatively low exciton binding energy, excitons in 3D OHP are easily dissociated into charge carriers which in turn decays through trap sites. Therefore, in this case, only longer-lived excitons ($\tau_1$) can contribute to PL intensity.[24] Some groups reported excess Pb atom induces non-radiative recombination or quenching in OHP materials.[11,23] Performed analyses were mainly focused on conversion of PbBr$_2$ to OHPs and grain size control after SVA. In SVA I, reduced metallic Pb atoms suppressed non-radiative recombination compared to pristine OHP film. The improvement of the recombination lifetime agrees with reduced PbBr$_2$ portion after SVA. From SVA I to SVA II, recrystallized OHP grains filled pin-hole and reduced grain boundaries which serve as non-radiative recombination. As non-radiative recombination is suppressed in SVA II, the longest exciton lifetime were observed. After the maximum luminance, excess vapour exposure resurrected sub-grain defects in SVA III and showed decreased radiative lifetime and portion, increased non-radiative portion. At last, enlarged OHP grain size increased non-radiative lifetime due to reduced possibility of encountering sub-grain defects.

**[Table 2]** Radiative and Non-radiative decay times and portions measured with TRPL. $\tau_1$ and $\tau_2$ indicates radiative and non-radiative recombination lifetimes (ns) and portions (%). The average exciton lifetime (ns) is written with $\tau_{av}$.

|  | Pristine | SVA I | SVA II | SVA III |
|---|---|---|---|---|
| $\tau_1$ | 3.109 ns | 17.553 ns | 35.297 ns | 22.975 ns |
|  | 44.84 % | 51.22 % | 87.80 % | 46.15 % |
| $\tau_2$ | 0.4129 ns | 1.4088 ns | 1.949 ns | 2.7912 ns |
|  | 55.16 % | 48.78 % | 12.20 % | 53.85 % |
| $\tau_{av}$ | 1.622 ns | 9.667 ns | 31.228 ns | 12.105 ns |

## Conclusion

In summary, we confirmed parameters of polar solvents to improve light emitting properties of organometallic halide perovskite polycrystalline thin films. Eventually, photoluminescence intensities of vapour treated films were significantly enhanced and the peak wavelengths were shifted to blue-colour region due to the formation of smaller grains by quantum confinement, from 530 nm to 480 nm. Exploited solvent vapour treatment for light emitters (LEDs and LDs) by this research which already exists in photovoltaic application was involved in extracting information about the correlation between grain size and photo-physical properties induced by various polar solvents due to their solubility and vapour pressure. Further researches will be focused on SVA effects in various halide perovskite materials and improving performances of optoelectronic devices.

## Acknowledgement

This research was supported by a National Research Foundation of Korea (NRF) grant funded by the Korean government Ministry of Science, ICT and Future Planning (MSIP) (NRF-2016R1A2B4013003) and (NRF-2015R1A2A1A01002493).

## References

1. Efficiency chart NREL, Solar Efficiency Chart https://www.nrel.gov/pv/assets/images/efficiency-chart.png (Accessed: March 2018)
2. A. Kojima, K. Teshima, Y. Shirai, and T. Miyasaka, *J. Am. Chem. Soc.,* 2009, **131**, 6050
3. H.-S. Kim, C.–R. Lim, J.-H. Im, K.-B. Lee, T. Moehl, A. Marchioro, S.-J. Moon, R. Humphry-Baker, J.-H. Yum, J. E. Moser, M. Grätzel and N.-G. Park, *Sci. Rep.*, 2012, **2**, 591
4. M. M. Lee, J. Teuscher, T. Miyasaka, T. N. Murakami and H. J. Snaith, *Science*, 2012, **338**, 643
5. J. H. Heo, D. H. Song, S. H. Im, *Adv. Mater.*, 2014, **26**, 8179
6. M. D. Xiao, F. Huang, W. Huang, Y. Dkhissi, Y. Zhu, J. Etheridge, A. Gray-Weale, U. Bach, Y.-B. Cheng, L. Spiccia, *Angew. Chem. Int. Ed.*, 2014, **53**, 9898
7. W. S. Yang, B. –W. Park, E. H. Jang, N. J. Jeon, Y. C. Kim, D. U. Lee, S. S. Shin, J. Seo, E. K. Kim, J. H. Noh and S. I. Seok, *Science*, 2017, **356**, 1376
8. S. –T. Ha, R. Su, J. Xing, Q. Zhang and Q. Xiong, *Chem. Sci.*, 2017, **8**, 2522
9. Z.-K. Tan, R. S. Moghaddam, M. L. Lai, P. Docampo, R. Higler, F. Deschler, M. Price, A. Sadhanala, L. M. Pazos, D. Credgington, F. Hanusch, T. Bein, H. J. Snaith and R. H. Friend, *Nat. Nanotech.*, 2014, **9**, 687
10. M. -G. La-Placa, G. Longo, A. Babaei, L. Martinez-Sarti, M. Sessolo and H. J. Bolink, *Chem. Commun.*, 2017, **53**, 8707
11. H. Cho, S.-H. Jeong, M.-H. Park, Y.-H. Kim, C. Wolf, C.-L. Lee, J. H. Heo, A. Sadhanala, N. S. Myoung, S. Yoo, S. H. Im, R. H. Friend and T.-W. Lee, *Science*, 2015, **350**, 1222
12. J. -W. Lee, Y. J. Choi, J. -M. Yang, S. Ham, S. K. Jeon, J. Y. Lee, Y. -H. Song, E. K. Ji, D. -H. Yoon, S. Seo, H. Shin, G. S. Han, H. S. Jung, D. Kim and N. -G. Park, *ACS Nano*, 2017, **11**, 3311
13. D. Li, S. A. Bretschneider, V. W. Bergmann, I. M. Hermes, J. Mars, A. Klasen, H. Lu, W. Tremel, M. Mezger, H.-J Butt, S. A. L. Weber and R. Berger, *J. Phys. Chem. C*, 2016, **120** (12), 6363
14. J. You, Y. M. Yang, Z. Hong, T.-B. Song, L. Meng, Y. Liu, C. Jiang, H. Zhou, W.-H. Chang, G. Li and Yang Yang, *Appl. Phys. Lett.*, 2014, **105**, 183902

15. G. E. Eperon, S. N. Habisreutinger, T. Leijtens, B. J. Bruijnaers, J. J. van Franeker, D. W. deQuilettes, S. Pathak, R. J. Sutton, G. Grancin, D. S. Ginger, R. A. J. Janssen, A. Petrozza and H. J. Snaith, *ACS Nano*, 2015, **9** (9), 9380
16. S. G. R. Bade, J. Li, X. Shan, Y. Ling, Y. Tian, T. Dilbeck, T. Besara, T. Geske, H. Gao, B. Ma, K. Hanson, T. Siegrist, C. Xu, and Z. Yu, *ACS Nano*, 2016, **10** (2), 1795
17. J. Huang, S. Tan, P. D. Lund and H. Zhou, *Energy Environ. Sci.,* 2017, **10**, 2284
18. Q. Chen, H. Zhou, Z. Hong, S. Luo, H.-S. Duan, H.-H. Wang, Y. Liu, G. Li and Yang Yang, *J. Am. Chem. Soc.*, 2014, **136**, 622
19. J. Liu, C. Gao, X. He, Q. Ye, L. Ouyang, D. Zhuang, C. Liao, J. Mei and W. Lau, *ACS Appl. Mater. Interfaces,* 2015, **7**, 24008
20. D. Di, K. P. Musselman, G. Li, A. Sadhanala, Y. Ievskaya, Q. Song, Z.-K. Tan, M. L. Lai, J. L. MacManus-Driscoll, N. C. Greenham and R. H. Friend, *J. Phys. Chem. Lett.,* 2015, **6**, 446
21. B. Delley, *J. Chem. Phys.,* 2000, **113**, 7756-7764
22. J. H. Kim and S. H. Kim, *Dyes Pigments,* 2016, **134**, 198
23. X. Fang, K. Zhang, Y. Li, L. Yao, Y. Zhang, Y. Wang, W. Zhai, L. Tao, H. Du and G. Ran. *Appl. Phys. Lett.,* 2016, **108**, 071109
24. D. Shi, V. Adinolfi, R. Comin, M. Yuan, E. Alarousu, A. Buin, Y. Chen, S. Hoogland, A. Rothenberger, K. Katsiev, Y. Losovyj, X. Zhang, P. A. Dowben, O. F. Mohammed, E. H. Sargent, O. M. Bakr, *Science,* 2015, **347**, 519
25. L. Zhang, X. Yang, Q. Jiang, P. Wang, Z. Yin, X. Zhang, H. Tan, Y. M. Yang, M. Wei, B. R. Sutherland, E. H. Sargent and J. You, *Nat. Commun.,* 2017, **8**, 15640

*Supplementary Information*

# Polar Solvent Vapour Induced Photo-Physical Properties on the Solid-State Organometallic Halide Perovskite Nanocrystals

Sang-Hyun Chin,[ab] Jin Woo Choi,[a] Hee Chul Woo,[a] Jong H. Kim,[c] Hong Seok Lee[*b] and Chang-Lyoul Lee[*a]

a. Advanced Photonics Research Institute (APRI), Gwangju Institute of Science and Technology (GIST), Gwangju, 61005, Republic of Korea. E-mail: jinwoo.choi@gist.ac.kr

b. Department of Physics, Research Institute of Physics and Chemistry (RINPAC), Chonbuk National University, Jeonju, 54896, Republic of Korea. E-mail: hslee1@jbnu.ac.kr

c. Department of Molecular Science and Technology, Ajou University, Suwon, 16499, Republic of Korea.

## I. Materials and Methods

### i. Thin film fabrication

All substrates were cleaned with detergent and sonicated in acetone, isopropanol (IPA), deionized water (DI Water) 15 minutes sequentially and washed with boiling isopropanol at last. After UV ozone treatment for 15 minutes, PEDOT:PSS was spun-coat on 3000 rpm 40 s and thermally annealed on 150 ℃ 20 min. 0.3 M Lead Bromide ($PbBr_2$ > 98% Aldrich, hereafter) was dissolved in N,N-Dimethylformamide (DMF) and spun-coat on substrate 3000 rpm 20 s.

Vapour-Assisted Solution Process (VASP) was waged right after $PbBr_2$ coating. 10 g of methylammonium bromide $CH_3NH_3Br$ (MABr) was sublimated at 125 ℃ 1.5 hours in closed glass petri dish (radius 45 mm, height 20 mm).

Demonstrated photoluminescence peak shift by quantum confinement effect and crystallinity enhancement was performed by solvent vapor annealing (SVA). The movie clip is attached. 100 ul of polar solvent (dimethylformamide, dimethylsulfoxide, deionized water, gamma-butylactone and ethanol) droplets for 8 points on petri dish (radius 30 mm, height 15 mm) was dropped and the dish was sealed right after.

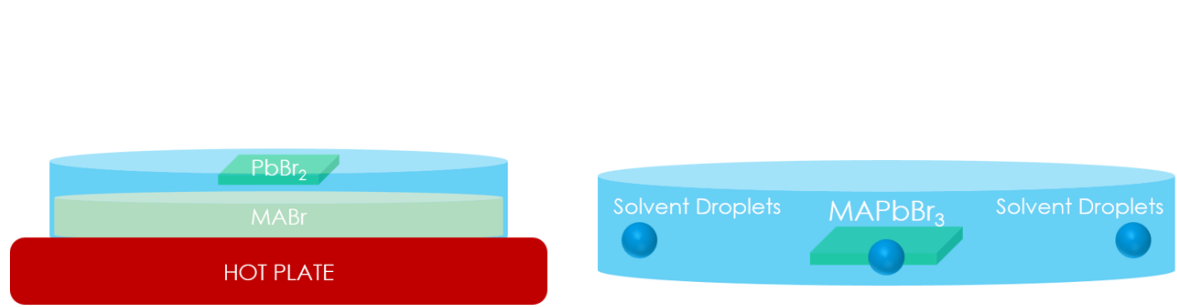

Schematic Illustration. Vapour-Assisted Solution Process & Solvent Vapour Annealing

### ii. XRD Measurements

The XRD measurements were performed on a Rigaku D/max-2500 diffractometer with Cu–Kα radiation (λ = 1.54 Å) at 40 kV and 100 mA.

### iii. Photoluminescence (PL) measurements

The steady-state PL of $MAPbBr_3$ layers on Glass/PEDOT:PSS/PMMA was measured by using a spectrofluorometer (JASCO FP6500).

### iv. Time-correlated single photon counting (TCSPC) measurements

The PL decay of $MAPbBr_3$ films on Glass/PEDOT:PSS/PMMA was investigated using a TCSPC system. A pulsed diode-laser head (LDH−P-C-405, PicoQuant) coupled with a laser-diode driver (PDL 800-B, PicoQuant) was used as an excitation source with a pulse width < 70 ps and a repetition rate of 5 MHz. The excitation wavelength was 405 nm. The fluorescence was spectrally resolved using a monochromator (SP-2150i, Acton) and its time-resolved signal was measured by a TCSPC module (PicoHarp, PicoQuant) with a microchannel plate photomultiplier tube (MCP-PMT, R3809U-59, Hamamatsu). The total instrument response function (IRF) was < 130 ps, and the temporal resolution was < 10 ps. The deconvolution of the decay curve, which separates the IRF and actual decay signal, was performed using fitting software (FluoFit, PicoQuant) to deduce the time constant associated with each exponential decay curve.

## II. DFT Calculation for various solvents

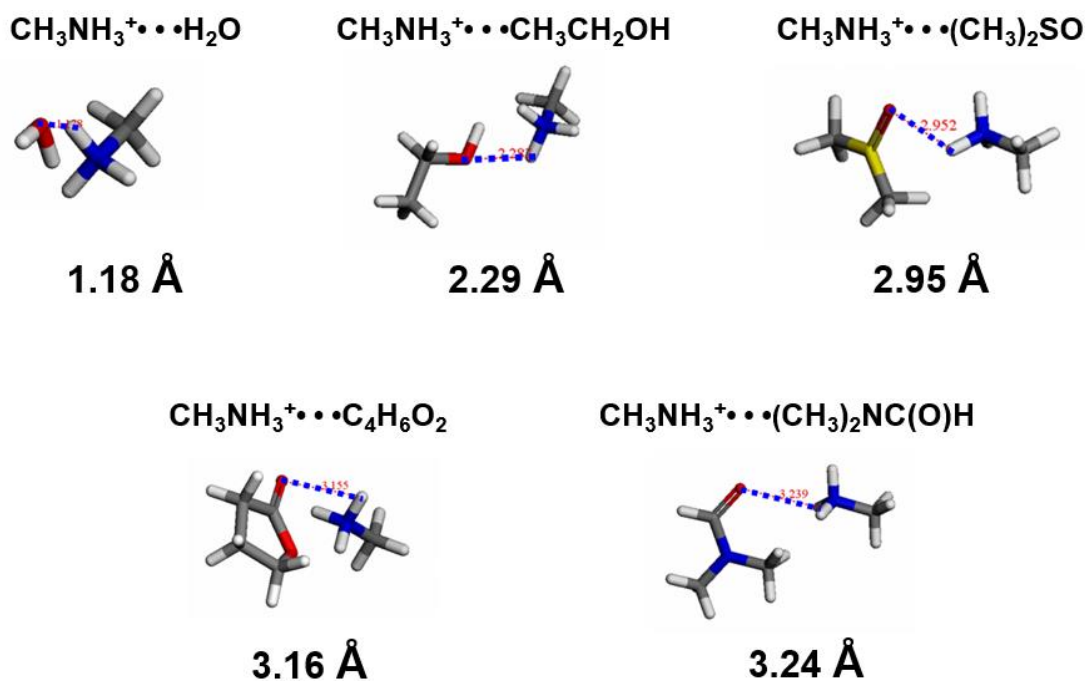

Figure S1. Interaction Distance between solvent molecules and methylammonium cation

|  | Ethanol | Water | DMF | GBL | DMSO |
|---|---|---|---|---|---|
| Chemical Formula | $CH_3CH_2OH$ | $H_2O$ | $(CH_3)_2NCOH$ | $C_4H_6O_2$ | $(CH_3)_2SO$ |
| Interaction Distance (Å) | 2.29 | 1.18 | 3.24 | 3.16 | 2.95 |
| Polarity (kJ/mol) | 51.9 | 63.1 | 43.8 | 44.3 | 45 |
| Vapor Pressure (mmHg) At 20℃ | 44.6 | 17.7 | 2.7 | 1.5 | 0.42 |

[1] J. H. Kim and S. H. Kim, *Dyes Pigments* 2016, **134**, 198-202 / **Interaction distance between Methylammonium cation & Polar solvent molecules**

[2] B. Delley, *J. Chem. Phys.* 2000, **113**, 7756-7764 / **DFT Calculation with DMol3**

## III. Photoluminescence spectrum with highly confined exciton

| | Analysis |
|---|---|
| PL Spectra | 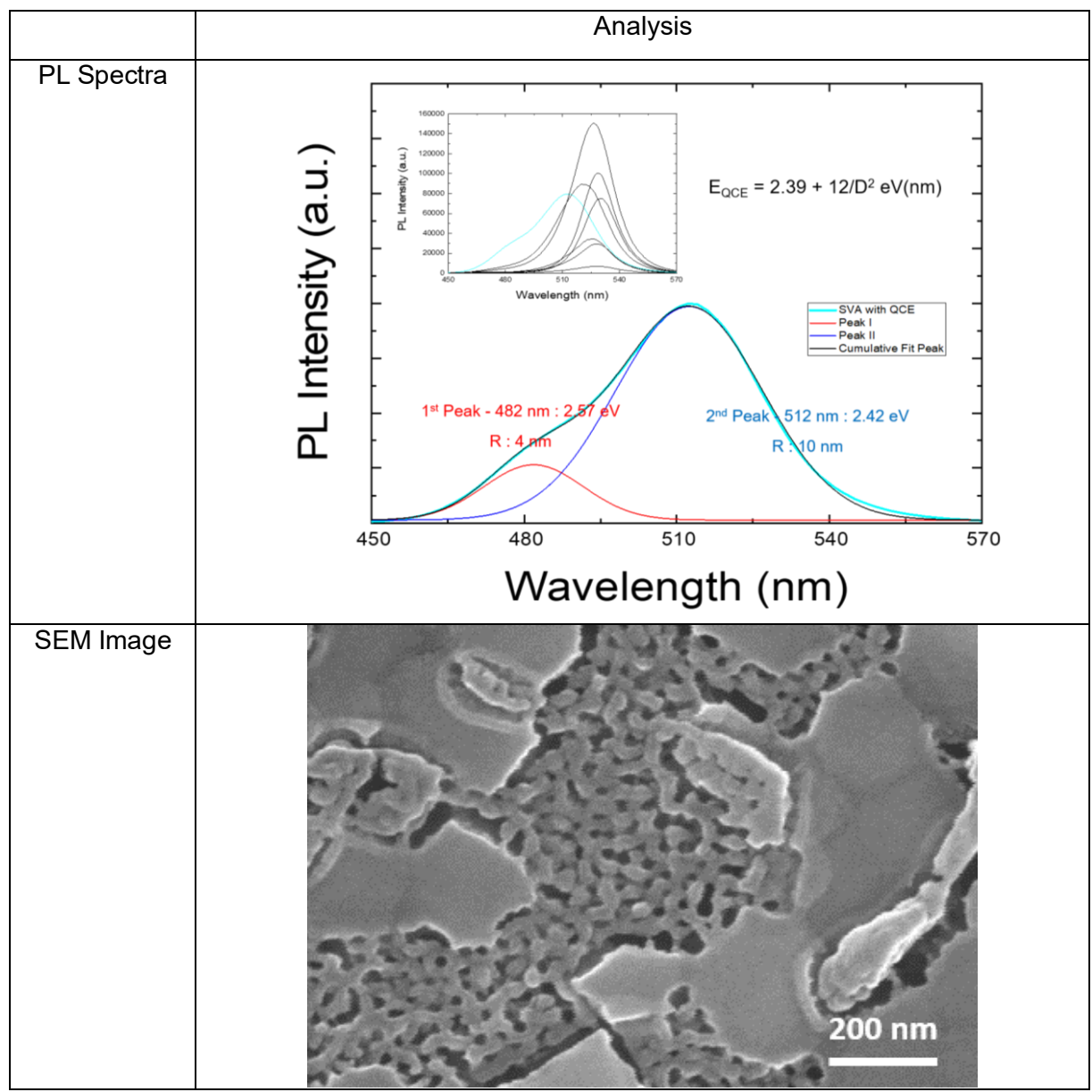 |
| SEM Image | |

Fig. S2 Gaussian analysis of leftmost PL spectrum and Perovskite grain

[3] D. Di, K. P. Musselman, G. Li, A. Sadhanala, Y. Ievskaya, Q. Song, Z.-K. Tan, M. L. Lai, J. L. MacManus-Driscoll, N. C. Greenham and R. H. Friend, *J. Phys. Chem. Lett.* 2015, **6**, 446−450 / **QCE Fitting Formula of Confined MAPbBr$_3$**

## IV. Scanning Electron Microscopy Images during Solvent Vapour Annealing

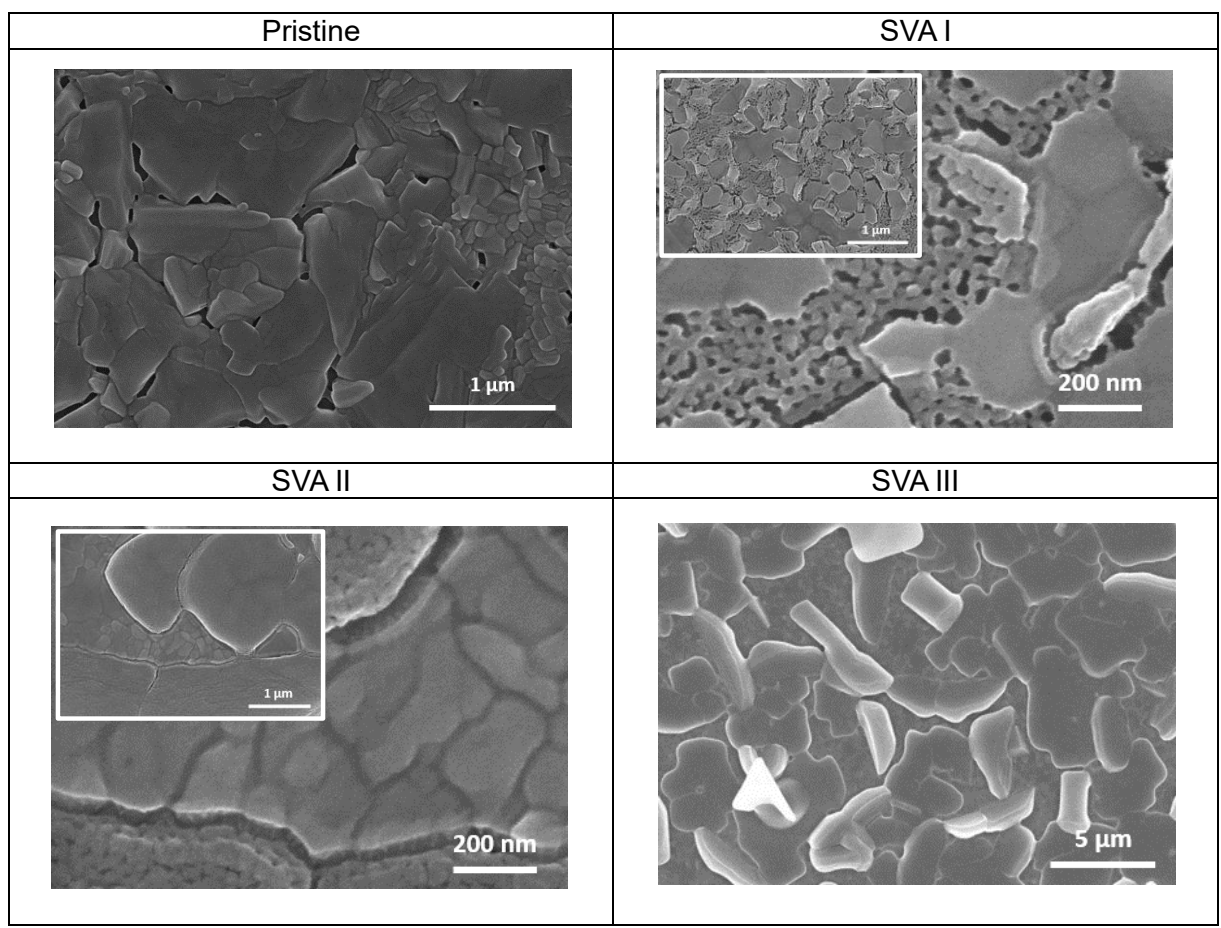

Fig. S3 SEM Images of Organometallic Halide Perovskite Grains during SVA

## *V. Macroscopic Solvent Vapour Annealing*

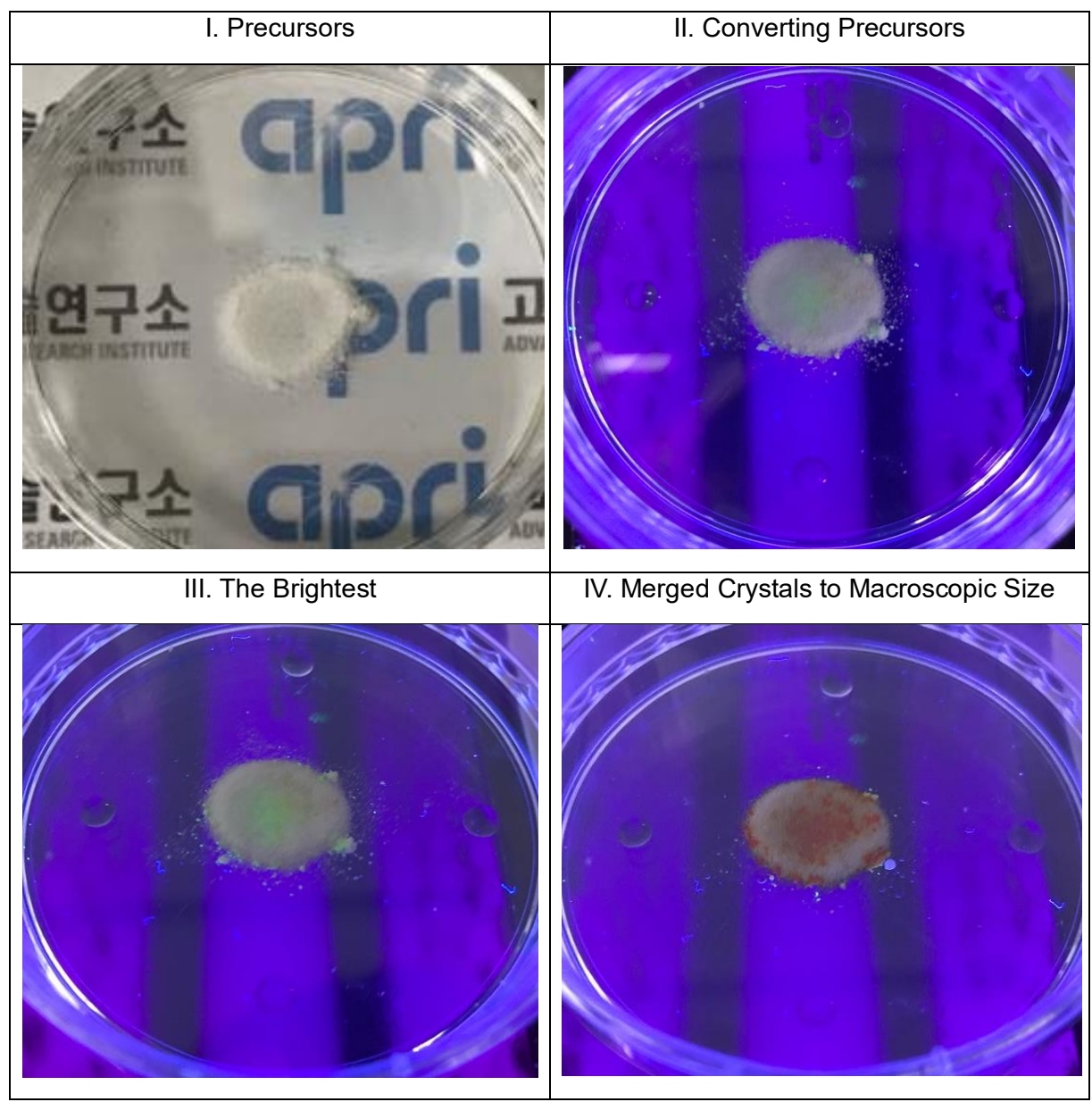

Fig. S4 $H_2O$ vapour converting precursors to perovskite crystals (I→IV)